\def\a{\alpha}
\def\b{\beta}
\def\c{\gamma}
\def\d{\delta}
\def\e{\epsilon}

\def\l{\lambda}
\def\m{\mu}
\def\n{\nu}

\def\r{\rho}
\def\s{\sigma}

\def\w{\omega}

\def\C{\Gamma}

\def\W{\Omega}

\def\pl{\partial}
\def\ul{\underline}
\def\rta{\rightarrow}

\documentclass{ws-procs975x65}

\begin{document}

\title{Causality and Superluminal Light}

\author{G.M. Shore}

\address{Department of Physics, University of Wales Swansea,\\
Singleton Park, Swansea SA2 8PP, U.K.\\
E-mail: g.m.shore@swansea.ac.uk}

\maketitle

\abstracts{The causal properties of curved spacetime, 
which underpin our sense of time in gravitational theories, are defined 
by the null cones of the spacetime metric. In classical general relativity, 
it is assumed that these coincide with the light cones determined by the
physical propagation of light rays. However, the quantum vacuum acts as
a dispersive medium for the propagation of light, since vacuum polarisation
in QED induces interactions which effectively violate the strong equivalence 
principle (SEP). For low frequencies the phenomenon of gravitational 
birefringence occurs and indeed, for some metrics and polarisations, photons 
may acquire {\it superluminal} phase velocities. In this article, we review 
some of the remarkable features of SEP violating superluminal propagation in 
curved spacetime and discuss recent progress on the issue of dispersion, 
explaining why it is the high-frequency limit of the phase velocity that 
determines the characteristics of the effective wave equation and thus the 
physical causal structure.}

\section{Introduction}

In general relativity, the nature of time and causality are determined by the 
properties of the null cones of the curved spacetime manifold. A physical
realisation of these geometric null cones is provided in classical electrodynamics
by the light cones traced by the propagation of light rays. As the title of this
conference implies, `Time and Matter' are therefore intimately related.

In quantum theory, however, the relationship is more subtle. Quantum effects
such as vacuum polarisation in QED induce interactions which effectively violate 
the strong equivalence principle and cause the quantum vacuum in the presence
of gravity to act as a dispersive medium for the propagation of light. At least
for low frequencies, the phenomenon of gravitational birefringence occurs and 
moreover, for some metrics and polarisations, photons may acquire {\it superluminal} 
phase velocities. This forces a reassessment of the identification of 
{\it light cones} with {\it null cones} and raises the question of how
causality and quantum theory can be reconciled in general relativity.

Research into this effect began with a key paper of Drummond and Hathrell\cite{DH} 
in which they calculated the one-loop vacuum polarisation 
contribution to the QED effective action in a background gravitational
field. They found the following modification to the free Maxwell action:
\begin{equation}
\begin{array}{rcl}
&\\
{}&\C = \int dx \sqrt{-g}\biggl[
-{1\over4}F_{\m\n}F^{\m\n} \\[4pt]
{}&+ {1\over m^2}\biggl(
a R F_{\m\n}F^{\m\n} + b R_{\m\n} F^{\m\l} F^\n{}_\l
+ c R_{\m\n\l\r} F^{\m\n} F^{\l\r} 
+ d D_\m F^{\m\l} D_\n F^\n{}_\l \biggr)\biggr] \\
&&
\end{array}\label{eq:aa}
\end{equation}
where $a$,$b$,$c$,$d$ are constants of $O(\a)$ and $m$ is the electron mass.
The important feature is the direct coupling of the electromagnetic field
to the curvature. This is an effective violation of the {\it strong} equivalence
principle (SEP), which is the dynamical ansatz that the `laws of physics' should be the 
same in the local inertial frames at each point in spacetime. (The {\it weak} equivalence
principle, viz.~the existence of such local inertial frames, is in contrast a
fundamental assumption underlying the structure of general relativity.)
More precisely, the SEP requires that electromagnetism is minimally coupled to gravity,
i.e.~through the connections only, independent of the curvature.
The effective action Eq.(\ref{eq:aa}) shows that while this principle may be 
consistently imposed at the classical level, it is necessarily violated in 
quantum electrodynamics. It is the quantum violation of the SEP that allows
the physical light cones to differ from the geometric null cones.

The new SEP-violating interactions in Eq.(\ref{eq:aa}) affect the propagation
of light and modify the physical light cones. Using the techniques of geometric
optics, Drummond and Hathrell showed that the new light cones are given by 
\begin{equation}
k^2 ~-~{2b\over m^2} R_{\m\l} k^\m k^\l ~+~{8c\over m^2} 
R_{\m\n\l\r} k^\m k^\l a^\n a^\r ~~=~~0
\label{eq:ab}
\end{equation}
where $k_\m$ is the wave vector and $a_\m$ the polarisation. Using the Einstein equations,
this can be expressed as the sum of a `matter' and a purely `gravitational' contribution
as follows:
\begin{equation}
k^2 ~-~{8\pi\over m^2}(2b+4c) T_{\m\l} k^\m k^\l ~+~{8c\over m^2} 
C_{\m\n\l\r} k^\m k^\l a^\n a^\r ~~=~~0
\label{eq:ac}
\end{equation}
The first term involves the projection of the energy-momentum tensor which appears
in the weak energy condition; this contribution is universal, appearing in the
modified light cone condition for light propagating subluminally in a variety of backgrounds
such as classical electromagnetic fields or finite temperature \cite{Sone,LPT,Sthree,Gies}. 
The second term depends on the Weyl tensor and is uniquely gravitational; since it depends
explicitly on the polarisation, the modified light cones exhibit {\it gravitational
birefringence}. 

It is also instructive to express Eqs.(\ref{eq:ab}),(\ref{eq:ac}) in the Newman-Penrose
formalism. Introducing a null tetrad with basis vectors  $\ell_\m$, $n_\m$, $m_\m$
and $\bar m_\m$ together with the corresponding components of the Ricci and Weyl
tensors $\Phi_{00}= -{1\over2}R_{\m\n}\ell^\m \ell^\n$, 
$\Psi_0 ~=~ - C_{\m\n\l\r}\ell^\m m^\n \ell^\l m^\r$ etc. 
(for details, see ref.\cite{Sthree}), 
and choosing $\ell_\m$ to coincide with the direction of propagation, i.e.
$k_\m = \omega \ell_\m$, we find
\begin{equation}
k^2 ~-~{(4b+8c)\omega^2\over m^2} \Phi_{00}  ~~\pm~~{4c\omega^2\over m^2} 
(\Psi_0 + \Psi_0^*) ~~=~~0
\label{eq:ad}
\end{equation}
This representation makes it clear that the contribution from the Weyl tensor changes
sign for the two physical transverse polarisations $a_\m$. It follows immediately that
for Ricci-flat spacetimes, both timelike and spacelike values of $k^2$ are possible.
In other words, Eqs.(\ref{eq:ab}),(\ref{eq:ac}),(\ref{eq:ad}) necessarily imply
the existence of {\it superluminal} propagation. Physical photons no
longer follow the geometrical null cones, but instead propagate on the
effective, polarisation-dependent, light cones defined above.

Many examples of the Drummond-Hathrell effect in a variety of gravitational
wave, black hole and cosmological spacetimes have been 
studied \cite{DH,Sone,Stwo,Sfour}.
The (Ricci-flat) black hole cases are particularly interesting, and it is
found that for photons propagating orbitally, the light cones are 
modified such that superluminal propagation occurs. For radial geodesics in
Schwarzschild spacetime, however, and the corresponding principal null geodesics
in Reissner-Nordstr\"om and Kerr, the light cones are unchanged. The reason is
simply that if we choose the standard Newman-Penrose tetrad in which
$\ell^\m$ is tangent to the principal null geodesic, the only non-vanishing component
of the Weyl tensor is $\Psi_2$ since these black hole spacetimes are all
Petrov type D, whereas the modification to the light cone condition
involves only $\Psi_0$. Superluminal propagation is also predicted in the
(Weyl-flat) FRW cosmological spacetimes, with the correction to the speed of light 
increasing as $1/t^2$ towards the initial singularity. Another interesting case involves
the Bondi-Sachs metric describing gravitational radiation from an isolated
source, where the magnitude of the superluminal effect is related to the
{\it peeling theorem} for the Weyl tensor. 

The existence of superluminal propagation in QED in curved spacetime of course raises
immediate questions as to the realisation of causality. The purpose of this 
article is to review our research programme on photon propagation in gravitational
fields with particular emphasis on the issue of causality and 
the consistency of quantum field theory with classical gravitation. We begin, 
in section 2, by considering carefully the implications of the Drummond-Hathrell
effective action as it stands, reviewing the {\it bimetric} interpretation of
the light cones, the realisation of {\it stable causality} in general relativity, 
possible time machine constructions and the consequences for
event horizons.

However, the action Eq.(\ref{eq:aa}) is only the lowest-order term
in an expansion of the full effective action in powers of derivatives. That is,
results derived from it are valid only in a low-frequency approximation
$\l \gg \l_c$, where $\l_c$ is the electron Compton wavelength. The
inclusion of terms of higher orders in derivatives shows that the propagation 
is in fact {\it dispersive}\cite{Sfive}. In section 3, we discuss the precise
definition of the `speed of light' and present a proof that the wavefront velocity, 
which is the relevant speed of light for causality, can be identified as the
high-frequency limit of the phase velocity, $v_{\rm ph}(\infty)$. This means that
a resolution of the issues raised in section 2 concerning causality 
depends on an explicit calculation of the light cones for {\it high-frequency}
propagation. In section 4, we present a recently derived extension of the
effective action valid to third order in curvatures and field strengths
and to all orders in derivatives\cite{Ssix}. The resulting light cone is derived
and some potential special features are described. Finally, however, we
argue on the basis of a detailed comparison with photon propagation in
background magnetic fields that a further, non-perturbative contribution
to the effective action may ultimately control the high-frequency limit
and we close with an assessment of the prospects for a final resolution
of the question of dispersion and causality for QED in curved spacetime.

\section{Causality and Superluminal Propagation}

In this section, we discuss the implications for causality of photon
propagation based on the Drummond-Hathrell action Eq.(\ref{eq:aa})
and its associated light cones, setting aside for the moment the issue
of dispersion. 

\subsection{Superluminal Propagation in Special and General Relativity}

It is generally understood that superluminal propagation in special relativity
leads to unacceptable violations of causality. Indeed the absence of tachyons
is traditionally employed as a constraint on fundamental theories. We therefore
begin by reviewing some basic features of superluminal propagation in 
order to sharpen these ideas in preparation for our subsequent discussion
of causality in general relativity.

The first important observation is that given a superluminal signal we can
always find a reference frame in which it is travelling backwards in time.
This is illustrated in Fig.~1. Suppose we send a signal from O to A at 
speed $v>1$ (in $c=1$ units) in frame ${\cal S}$ with coordinates $(t,x)$.
In a frame ${\cal S}'$ moving with respect to ${\cal S}$ with velocity 
$u>{1\over v}$, the signal travels backwards in $t'$ time, as follows
immediately from the Lorentz transformation.\footnote{From 
the Lorentz transformations,
we have $t'_A = \c(u)t_A (1-uv)$ and $x'_A = \c(u)x_A (1-{u\over v})$
For the situation realised in Fig.~1, we require both $x'_A >0$ and $t'_A <0$,
that is ${1\over v} < u < v$, 
which admits a solution only if $v>1$.} 

\begin{figure}[t]
{\epsfxsize=11cm\epsfbox{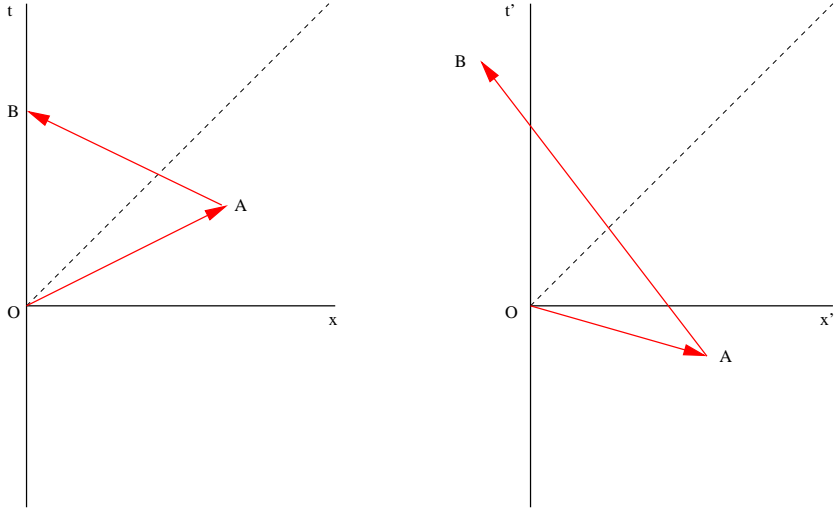}}
\caption{A superluminal $(v>1)$ signal OA which is forwards 
in time in frame ${\cal S}$ is backwards in time in a frame ${\cal S}'$ moving
relative to ${\cal S}$ with speed $u>{1\over v}$. However, the return path with
the same speed in ${\cal S}$ arrives at B in the future light cone of O, 
independent of the frame.  \label{fig:1}}
\end{figure}

The important point for our considerations is that this {\it by itself} 
does not necessarily imply a violation of causality. For this, we require that 
the signal can be returned from A to a point in the past light cone of O. 
However, if we return the signal from A to B with the same speed in frame ${\cal S}$,
then of course it arrives at B in the future cone of O. The situation is physically
equivalent in the Lorentz boosted frame ${\cal S}'$ -- the return signal travels 
forward in $t'$ time and arrives at B in the future cone of O. This, unlike
the assignment of spacetime coordinates, is a frame-independent statement.

The problem with causality arises from the scenario illustrated in Fig.~2.
\begin{figure}[t]
{\epsfxsize=5cm\epsfbox{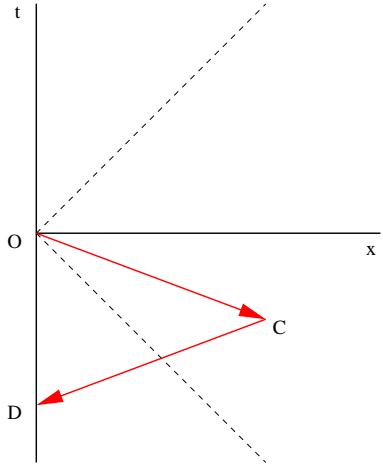}}
\caption{A superluminal $(v>1)$ signal OC which is backwards 
in time in frame ${\cal S}$ is returned at the same speed to point D in the past light
cone of O, creating a closed time loop.  \label{fig:2}}
\end{figure}
Clearly, if a backwards-in-time signal OC is possible in frame ${\cal S}$,
then a return signal sent with the same speed will arrive at D in the past light
cone of O creating a closed time loop OCDO.
The crucial point is that local Lorentz invariance of the laws 
of motion implies that if a superluminal signal such as OA is possible, then so is one
of type OC, since it is just given by an appropriate Lorentz boost (as in Fig.~1).
The existence of {\it global} inertial frames then guarantees the existence of
the return signal CD (in contrast to the situation in Fig.~1 viewed in the
${\cal S}'$ frame). 

The moral is that {\it both} conditions must be met in order to guarantee the 
occurrence of unacceptable closed time loops -- the existence of a superluminal signal
{\it and} global Lorentz invariance. Of course, since global Lorentz invariance
(the existence of global inertial frames) is the essential part of the structure
of special relativity, we recover the conventional wisdom that in this theory, 
superluminal propagation is indeed in conflict with causality.\footnote{Notice that this 
is not in contradiction with the occurrence of superluminal propagation in flat
spacetime with Casimir plates\cite{Scharn,Barton,LSV},
since the modification of the spacetime geometry by the plates removes {\it global}
Lorentz invariance.} 

The reason for presenting this elementary discussion is to emphasise that the situation
is crucially different in general relativity. The {\it weak} equivalence principle, 
which we understand as the statement that a {\it local} inertial 
frame exists at each point in spacetime, implies that general relativity is
formulated on a Riemannian manifold. However, local Lorentz invariance alone is not
sufficient to establish the link between superluminal propagation
and causality violation. This is usually established by adding 
a second, dynamical, assumption. The {\it strong}
equivalence principle (SEP) states that the laws of physics should be identical in
the local frames at different points in spacetime, and that they should
reduce to their special relativistic forms at the origin of each local frame.
It is the SEP which takes over the role of the existence of global inertial frames
in special relativity in establishing the incompatibility of superluminal
propagation and causality.

However, unlike the weak equivalence principle, which underpins the essential
structure of general relativity, the SEP is merely a simplifying
assumption about the dynamics of matter coupled to gravitational fields.
Mathematically, it requires that matter or electromagnetism is {\it minimally
coupled} to gravity, i.e.~with interactions depending only on the connections
but not the local curvature. This ensures that at the origin of a local frame,
where the connections may be Lorentz transformed locally to zero, the dynamical
equations recover their special relativistic form. In particular, the SEP is
violated by interactions which explicitly involve the curvature, such as
those occurring in the Drummond-Hathrell action (\ref{eq:aa}) and the consequent
modified light cones.

The question of whether this specific realisation of superluminal propagation is
in conflict with causality is discussed in section 2.4 using the concept
of {\it stabele causality} described in ref.\cite{HE}. Notice though that by
violating the SEP, we have evaded the {\it necessary} association of superluminal
motion with causality violation that held in special relativity. Referring back to
the figures, what is established is the existence of a signal of type OA,
which as we saw, does not by itself imply problems with causality even though frames 
such as ${\cal S}'$ exist {\it locally} with respect to which motion is backwards in 
time. However, since the SEP is broken, even if a local frame exists in which the 
signal looks like OC, it does {\it not} follow that a return path CD is allowed. 
The signal propagation is fixed, determined locally by the spacetime curvature.

\subsection{Geometric Optics}

The most direct way to deduce the form of the light cones for QED in curved spacetime
is to use geometric optics. This starts from the ansatz 
\begin{equation}
(A_\m + i\e B_\m + \ldots ) \exp\bigl(i {\vartheta\over\e}\bigr)
\label{eq:ba}
\end{equation}
in which the electromagnetic field is written as a slowly-varying amplitude and
a rapidly-varying phase. The parameter $\e$ is introduced as a device to keep
track of the relative order of magnitude of terms, and the Bianchi and 
Maxwell equations are solved order-by-order in $\e$.

The wave vector is identified as the gradient of the phase, 
$k_\m = \partial_\m\vartheta$. 
We also write $A_\m = A a_\m$, where $A$ represents
the amplitude itself while $a_\m$ specifies the polarisation, which
satisfies $k_\m a^\m = 0$. 

Solving the usual Maxwell equation $D_\m F^{\m\n} = 0$, we find at $O(1/\e)$,
\begin{equation}
k^2 = 0
\label{eq:bb}
\end{equation}
while at $O(1)$,
\begin{equation}
k^\m D_\m a^\n = 0
\label{eq:bc}
\end{equation}
and
\begin{equation}
k^\m D_\m(\ln A) = -{1\over2}D_\m k^\m
\label{eq:bd}
\end{equation}
Eq.(\ref{eq:bb}) shows immediately that $k^\m$ is a null vector. From its
definition as a gradient, we also see
\begin{equation}
k^\m D_\m k^\n  = k^\m D^\n k_\m = {1\over2}D^\n k^2 = 0
\label{eq:be}
\end{equation}
Light rays, or equivalently photon trajectories, are the integral curves of 
$k^\m$, i.e.~the curves $x^\m(s)$ where $dx^\m/ds = k^\m$. These curves 
therefore satisfy
\begin{equation}
0 ~=~ k^\m D_\m k^\n 
~=~ {d^2 x^\n \over ds^2} + \C^\n_{\m\l} {dx^\m \over ds}{dx^\l \over ds}
\label{eq:bf}
\end{equation}
This is the geodesic equation. We conclude that for the usual Maxwell theory
in general relativity, light follows null geodesics. 
Eqs.(\ref{eq:be}),(\ref{eq:bc}) show that both the wave vector and the polarisation 
are parallel transported along these null geodesic rays, while Eq.(\ref{eq:bd}),
whose r.h.s.~is just (minus) the optical scalar $\theta$, shows
how the amplitude changes as the beam of rays focuses or diverges.

\subsection{Bimetricity}

The same method is applied to the modified Maxwell equation derived from the
effective action Eq.(\ref{eq:aa}):
\begin{equation}
D_\m F^{\m\n} ~-~ {1\over m^2} \biggl[2b R_{\m\l}D^\m F^{\l\n}
+ 4c R_\m{}^\n{}_{\l\r}D^\m F^{\l\r}\biggr]~~=~~0
\label{eq:bg}
\end{equation}
from which the new light cone condition 
\begin{equation}
k^2 ~-~{2b\over m^2} R_{\m\l} k^\m k^\l ~+~{8c\over m^2} 
R_{\m\n\l\r} k^\m k^\l a^\n a^\r ~~=~~0
\label{eq:bh}
\end{equation}
follows immediately. Since this new light cone relation is
still homogeneous and quadratic in $k^\m$, we can write it as
\begin{equation}
{\cal G}^{\m\n}k_\m k_\n = 0
\label{eq:bi}
\end{equation}
defining ${\cal G}^{\m\n}$ as the appropriate function of the curvature
and polarisation. 

Now notice that in the discussion of the free Maxwell theory, we did not need to
distinguish between the photon momentum  $p^\m$, i.e.~the tangent vector 
to the light rays, and the wave vector $k_\m$ since they were simply related
by raising the index using the spacetime metric, $p^\m = g^{\m\n}k_\n$. 
In the modified theory, however, there is an important distinction. 
The wave vector, defined as the gradient of the phase, is a covariant vector 
or 1-form, whereas the photon momentum/tangent vector to the rays is a true 
contravariant vector. The relation is non-trivial. 
In fact, given $k_\m$, we should define the corresponding
`momentum' as 
\begin{equation}
p^\m = {\cal G}^{\m\n}k_\n
\label{eq:bj}
\end{equation}
and the light rays as curves $x^\m(s)$ where ${dx^\m\over ds} = p^\m$.
This definition of momentum satisfies 
\begin{equation}
G_{\m\n} p^\m p^\n = {\cal G}^{\m\n}k_\m k_\n = 0
\label{eq:bk}
\end{equation}
where $G \equiv {\cal G}^{-1}$ defines a new {\it effective metric}
which determines the light cones mapped out by the geometric optics light rays.
(Indices are always raised or lowered using the true metric $g_{\m\n}$.)
The {\it ray velocity} $v_{\rm ray}$ corresponding to the momentum $p^\m$, 
which is the velocity with which the equal-phase surfaces advance, is given by
(defining components in an orthonormal frame)
\begin{equation}
v_{\rm ray} = {|\ul p|\over p^0} = {d |\ul x|\over dt}
\label{eq:bl}
\end{equation}
along the ray. This is in general different from the {\it phase velocity}
\begin{equation}
v_{\rm ph} = {k^0\over|\ul k|}
\label{eq:bm}
\end{equation}

This shows that photon propagation for QED in curved spacetime can be
characterised as a {\it bimetric} theory\footnote{
Bimetric theories of gravity have an extensive literature. See ref.\cite{Drum}
for an elegant recent construction and references therein for earlier
work.} -- the physical light cones are
determined by the effective metric $G_{\m\n}$ while the geometric null
cones are fixed by the spacetime metric $g_{\m\n}$.

\subsection{Stable Causality}

The bimetric formulation is the most natural language in which to discuss whether
the superluminal velocities predicted by the Drummond-Hathrell action
are compatible with our usual idea of causality. 
We have already seen that with SEP violation in general relativity,
the arguments that in special relativity led to the incompatibility of superluminal
propagation and causality are no longer valid. Superluminal motion {\it may} 
be possible -- the question is to find a criterion to decide 
whether it {\it is}\footnote{The 
equivalent discussion of causality for superluminal propagation
in Minkowski spacetime with Casimir plates is given in ref.\cite{LSV}. 
This also provides a nice example of the distinction between
ray and phase velocities discussed above.}.

One special case where causality is realised in a particularly simple way is in
globally hyperbolic spacetimes, where the manifold admits a foliation 
into a set of spacelike Cauchy surfaces with fibres given by timelike geodesics.
It is not hard to imagine that the same structure could be preserved using the 
effective metric $G_{\m\n}$ to define `spacelike' or `timelike', especially if
$G_{\m\n}$ is only perturbatively different from the actual spacetime metric $g_{\m\n}$.
But this would be a global question and the preservation of global hyperbolicity
is not {\it a priori} guaranteed. 

The clearest criterion for causality in general involves the concept of 
{\it stable causality} discussed, for example, in the monograph of Hawking and 
Ellis\cite{HE}. Proposition 6.4.9 states the required
definition and theorem:

\noindent $\bullet$ A spacetime manifold $({\cal M},g_{\m\n})$ is {\it stably causal} 
if the metric $g_{\m\n}$ has an open neighbourhood such that ${\cal M}$ has no closed 
timelike or null curves with respect to any metric belonging to that 
neighbourhood.

\noindent $\bullet$ Stable causality holds everywhere on ${\cal M}$ if and only if there
is a globally defined function $f$ whose gradient $D_\m f$ is everywhere non-zero
and timelike with respect to $g_{\m\n}$.  

According to this theorem, the absence of causality violation in the form of closed
timelike or lightlike curves is assured if we can find a globally defined function $f$ whose
gradient is timelike {\it with respect to the effective metric} $G_{\m\n}$ for light
propagation. $f$ then acts as a global time coordinate. 

To see how this criterion can be applied to a particular example, one for which
stable causality {\it is} preserved by the new light cone metric, consider the
cosmological Friedmann-Robertson-Walker spacetime. 
Since the FRW metric is Weyl flat, the modified light cone condition Eq.(\ref{eq:ac})
reads simply
\begin{equation}
k^2 = \zeta~ T_{\m\n} k^\m k^\n 
\label{eq:bn}
\end{equation}
where $\zeta = {8\pi\over m^2}(2b+4c)$ and the energy-momentum tensor is
\begin{equation}
T_{\m\n} = (\r + P)n_\m n_\n - P g_{\m\n}
\label{eq:bo}
\end{equation}
with $n^\m$ specifying the time direction in a comoving
orthonormal frame. $\r$ is the energy density and $P$ is the
pressure, which in a radiation-dominated era are related by $\r - 3P = 0$.
The phase velocity is independent of polarisation and is found to be
superluminal\footnote{In the radiation dominated era, 
where $\r(t) = {3\over 32\pi} t^{-2}$, we have
\begin{equation}
v_{\rm ph} = 1 + {1\over 16\pi}\zeta~ t^{-2}
\end{equation}
which, as already observed in ref.\cite{DH}, increases towards the early
universe. Although this expression is only reliable in the perturbative
regime where the correction term is small, it is intriguing that QED
predicts a rise in the speed of light in the early universe. It is interesting
to speculate whether this superluminal effect persists for high curvatures near 
the initial singularity and whether it could play a role in resolving
the horizon problem in cosmology}:
\begin{equation}
v_{\rm ph} = {k^0\over |\ul k|} = 1 + {1\over2}\zeta (\r + P)
\label{eq:bp}
\end{equation}
At first sight, this looks surprising given that $k^2 > 0$, its sign
fixed by the weak energy condition $T_{\m\n} k^\m k^\n \ge 0$.
However, if instead we consider the momentum along the rays,
$p^\m = {\cal G}^{\m\n}k_\n$,
we find
\begin{equation}
p^2 = g_{\m\n}p^\m p^\n = -\zeta (\r + P) (p^0)^2
\label{eq:bq}
\end{equation}
and 
\begin{equation}
v_{\rm ray} = {|\ul p|\over p^0} = 1 + {1\over2}\zeta(\r + P)
\label{eq:br}
\end{equation}
The effective metric $G = {\cal G}^{-1}$ is (in the orthonormal frame)
\begin{equation}
G ~=~ \left(\matrix{1 + \zeta\r &0&0&0\cr
0&-(1-\zeta P)&0&0\cr 0&0&-(1-\zeta P)&0\cr 0&0&0&-(1-\zeta P)\cr}\right)
\label{eq:bs}
\end{equation}
In this case, therefore, we find equal and superluminal
velocities $v_{\rm ph} = v_{\rm ray}$ and $p^2 < 0$ is manifestly
spacelike as required. 

Is stable causality preserved? In this case the answer is yes, since
we may still use the cosmological time coordinate $t$ as the globally
defined function $f$. We need only check that $D_\m t$ defines a 
timelike vector with respect to the effective metric $G_{\m\n}$. 
This is true provided $G_{00} > 0$, which is certainly satisfied
by Eq.(\ref{eq:bs}). So at least in this case, superluminal propagation 
is compatible with causality.

\subsection{Time Machines?}

Although we have seen that causality is not necessarily
violated by superluminal propagation, it is important to look for 
counter-examples where the Drummond-Hathrell effect may create a
time machine. One imaginitive suggestion was put forward by Dolgov
and Novikov (DN)\cite{DN}, involving two gravitating sources in relative motion.
This scenario therefore has some echoes of the Gott cosmic string time
machine\cite{Gott}; both are reviewed in ref.\cite{Sseven}.

The DN proposal is to consider first a gravitating source with a superluminal
photon following a trajectory which we may take to be radial. Along this 
trajectory, the metric interval is effectively two-dimensional and DN
consider the form
\begin{equation}
ds^2 =  A^2(r) dt^2 - B^2(r) dr^2
\label{eq:bt}
\end{equation}
(An explicit realisation is given by radial superluminal signals in the Bondi-Sachs
spacetime, described in ref.\cite{Sfour}.) 
The photon velocity in the $(t,r)$ coordinates
is taken to be $v = 1 + \d v$, so the effective light cones lie perturbatively
close to the geometric ones. The trajectory is forward in time with respect 
to $t$.

DN now make a coordinate transformation corresponding to a frame in relative
motion to the gravitating source, rewriting the metric interval along the
trajectory as
\begin{equation}
ds^2 = A^2(t',r') \bigl(dt'^2 - dr'^2\bigr)
\label{eq:bu}
\end{equation}
The transformation is\footnote{This transformation comprises two steps.
First, since any 2-dim metric is conformally flat, we can bring the
metric Eq.(\ref{eq:bt}) into standard form $ds^2 = \Omega^2
\bigl(d\tilde t^2 - d\tilde r^2\bigr)$. Then, a boost with velocity $u$ is made
on the flat coordinates $(\tilde t, \tilde r)$ to give the DN coordinates
$(t',r')$.}
\begin{equation}
\begin{array}{rcl}
&\\
t' &= \c(u)\bigl(t - ur -u f(r) \bigr) \\[4pt]
r' &= \c(u)\bigl(r - ut + f(r) \bigr) \\
&&
\end{array}\label{eq:bv}
\end{equation}
with 
\begin{equation}
f(r) = \int dr~\Bigl({B\over A} - 1\Bigr)
\label{eq:bw}
\end{equation}

Now, a superluminal signal with velocity 
\begin{equation}
v = 1 + \d v = {B\over A} {dr\over dt}
\label{eq:bz}
\end{equation}
emitted at $(t_1,r_1)$ and received at $(t_2,r_2)$ travels forward in $t$
time (for small, positive $\d v$) with interval
\begin{equation}
t_2 - t_1 = \int_{r_1}^{r_2} dr ~\bigl(1- \d v\bigr) {B\over A}
\label{eq:bza}
\end{equation}
As DN show, however, this motion is backwards in $t'$ time for sufficiently
large $u$, since the equivalent interval is
\begin{equation}
t'_2 - t'_1 = \c(u)~\int_{r_1}^{r_2} dr ~\bigl(1- u -\d v\bigr) {B\over A}
\label{eq:bzb}
\end{equation}
The required frame velocity is $u > 1 - \d v$, i.e.~since $\d v$ is small,
$u > {1\over v}$.

The situation so far is therefore identical in principle to the discussion of 
superluminal propagation illustrated in Fig.~1. In DN coordinates 
the outward superluminal signal is certainly propagating backwards in time, but a 
reverse path with the same perturbatively superluminal velocity would necessarily
go sufficiently forwards in time to arrive back within the future light cone
of the emitter.

At this point, however,
DN propose to introduce a second gravitating source moving relative to the
first, as illustrated in Fig.~3.
\begin{figure}[t]
{\epsfxsize=9cm\epsfbox{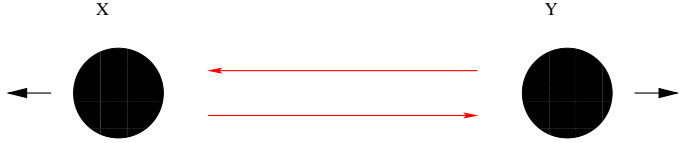}} 
\caption{The Dolgov-Novikov time machine proposal. A
superluminal signal from $X$, described as backwards-in-time in a relevant 
frame, is sent towards a second gravitating source $Y$ moving
relative to $X$ and returned symmetrically.  \label{fig:3}}
\end{figure}
They now claim that a superluminal photon emitted with velocity $v(r)$ in the
region of $X$ will travel backwards in time (according to the physically
relevant coordinate $t'$) to a receiver in the region of $Y$. A signal
is then returned symmetrically
to be received at its original position in the vicinity of $X$, arriving,
according to DN, in its past. This would then be analogous to the 
situation illustrated in Fig.~2.  

However, as we emphasised in section 2.1, we are {\it not} free to realise the
scenario of Fig.~2 in the gravitational case, because the SEP-violating 
superluminal propagation proposed by Drummond and Hathrell is pre-determined, 
fixed by the local curvature. The $t'$ frame may describe back-in-time 
motion for the outward leg, but it does not follow that the return path
is similarly back-in-time {\it in the same frame.} The appropriate special 
relativistic analogue is the scenario of Fig.~1, not Fig.~2. This critique of 
the DN time machine proposal has already been made by Konstantinov\cite{Konst} 
and further discussion of the related effect in flat spacetime with Casimir 
plates is given in ref.\cite{LSV}. The relative motion of the two sources, which 
at first sight seems to allow the backwards-in-time coordinate $t'$ to be 
relevant and to be used symmetrically, does not in fact alleviate the problem.
 
The true situation seems rather more to resemble Fig.~4.
\begin{figure}[t]
{\epsfxsize=9cm\epsfbox{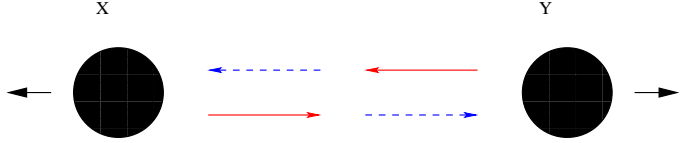}}
\caption{A decomposition of the paths in Fig.~3 for
well-separated sources  \label{fig:4}}
\end{figure}
With the gravitating sources $X$ and $Y$ sufficiently distant that spacetime
is separated into regions where it is permissible to neglect one or the other,
a signal sent from the vicinity of $X$ towards $Y$ and back would follow
the paths shown. But it is clear that this is no more than stitching together
an outward plus inward leg near source $X$ with an inward plus outward leg
near $Y$. Since both of these are future-directed motions, in the sense of
Fig.~1, their combination cannot produce a causality-violating trajectory.
If, on the other hand, we consider $X$ and $Y$ to be sufficiently close that
this picture breaks down, we lose our ability to analyse the Drummond-Hathrell
effect, since we would need the full collision metric for the gravitating
sources which is not known for physically realisable examples. 

We therefore conclude that the Dolgov-Novikov time machine does not work. 
The essential idea
of trying to realise the causality-violating special relativistic scenario 
of Fig.~2 by using two gravitational sources in relative motion does not 
in the end succeed, precisely because the physical Drummond-Hathrell light cones
are fixed by the local curvature. Once more it appears that in general relativity 
with SEP-violating interactions, superluminal photon propagation 
and causality can be compatible.

\subsection{The Event Horizon}

We have seen that when quantum effects are taken into account, the physical
light cones need not coincide with the geometrical null cones.
This immediately raises the question of black hole event horizons --
do the physical horizons for light propagation also differ from the geometrical
horizons, and are they polarisation dependent? If so, this would have 
profound repercussions for phenomena such as Hawking radiation.

The answer is best seen using the Newman-Penrose form of the light cone,
viz.
\begin{equation}
k^2 ~-~{(4b+8c)\omega^2\over m^2} \Phi_{00}  ~~\pm~~{4c\omega^2\over m^2} 
(\Psi_0 + \Psi_0^*) ~~=~~0
\label{eq:bzc}
\end{equation}
If we define the tetrad with $\ell^\m$ as an outward-directed null vector
orthogonal to the horizon 2-surface, then a fundamental theorem on horizons
states that both $\Phi_{00}$ and $\Psi_0$ are zero precisely at the 
horizon. The detailed proof, which is
given in ref.\cite{Hawk}, involves following the convergence and shear
of the generators of the horizon. In physical terms, however, it is easily
understood as the requirement that the flow of both matter (given by the
Ricci term) and gravitational radiation (given by the Weyl term) are zero
across the horizon.

It follows that for outward-directed photons with $k_\m = \omega \ell_\m$, 
the quantum corrections vanish at the horizon and the
light cone coincides with the null cone. The geometrical event horizon is 
indeed the true horizon for physical photon propagation\cite{Sthree,Gibb}. 
Again, no conflict arises between superluminal propagation and essential causal 
properties of spacetime.

\section{Causality, Characteristics and the `Speeds of Light'}

So far, our analysis of photon propagation has been based entirely on the
leading-order, Drummond-Hathrell effective action Eq.(\ref{eq:aa}).
However, as we show in section 4, the full effective action contains terms to
all orders in a derivative expansion and these must be taken into account to go
beyond the low-frequency approximation. Photon propagation in QED in curved
spacetime is therefore dispersive and we must understand how to identify
the `speed of light' which is relevant for causality.

\subsection{`Speeds of Light'}

An illuminating discussion of wave propagation in a simple
dispersive medium is given in the classic work by Brillouin\cite{Brill}. 
This considers propagation of a sharp-fronted pulse of waves in a medium 
with a single absorption band, with refractive index $n(\w)$:
\begin{equation}
n^2(\w) = 1 - {a^2\over \w^2 -\w_0^2 + 2i\w\r}
\label{eq:ca}
\end{equation}
where $a,\r$ are constants and $\w_0$ is the characteristic frequency
of the medium. Five\footnote{In fact, if we take into account the distinction
discussed in section 2 between the {\it phase velocity} $v_{\rm ph}$ and the 
{\it ray velocity} $v_{\rm ray}$, and include the fundamental speed of light 
constant $c$ from the Lorentz transformations, we arrive at {\it seven} 
distinct definitions of `speed of light'.}
distinct velocities are identified: the {\it phase
velocity} $v_{\rm ph} = {\w\over{|\ul k|}} = \Re{1\over n(\w)}$, 
{\it group velocity} $v_{\rm gp} = {d\w\over d|\ul k|}$,
{\it signal velocity} $v_{\rm sig}$, {\it energy-transfer velocity} 
$v_{\rm en}$ and {\it wavefront velocity} $v_{\rm wf}$, with precise 
definitions related to the behaviour of contours and saddle points in the
relevant Fourier integrals in the complex $\w$-plane.
Their frequency dependence is illustrated in Fig.~5.
\begin{figure}[t]
{\epsfxsize=8cm\epsfbox{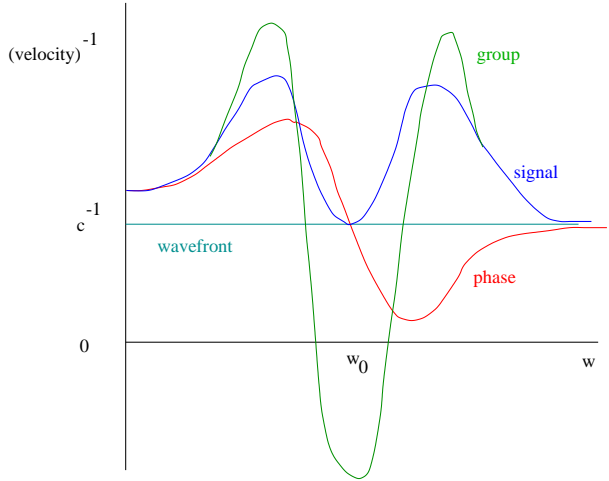}}
\caption{ Sketch of the behaviour of the phase, group and 
signal velocities with frequency in the model described by the refractive
index Eq.(\ref{eq:ca}). The energy-transfer velocity (not shown) is always 
less than $c$ and becomes small near $\w_0$. The wavefront speed is 
identically equal to $c$. \label{fig:5}}
\end{figure}
As the pulse propagates, the first disturbances to arrive are
very small amplitude waves, `frontrunners', which define the wavefront
velocity $v_{\rm wf}$. These are followed continuously by waves with amplitudes
comparable to the initial pulse; the arrival of this part of the complete
waveform is identified in ref.\cite{Brill} as the signal velocity $v_{\rm sig}$.
As can be seen from Fig.~5, it essentially coincides with the more familiar
group velocity for frequencies far from $\w_0$, but gives a much more
intuitively reasonable sense of the propagation of a signal than the group 
velocity, whose behaviour in the vicinity of an absorption band is
relatively eccentric.\footnote{Notice that it is the group velocity which is 
measured in quantum optics experiments which find light speeds of
essentially zero\cite{Hau} or many times $c$ \cite{WKD}. A particularly
clear description in terms of the effective refractive index is given
in ref.\cite{Hau}.}
As the figure makes clear, the phase velocity itself also does not
represent a `speed of light' relevant for considerations of signal
propagation or causality.

The appropriate velocity to define light cones and causality is in fact
the wavefront velocity $v_{\rm wf}$. (Notice that in Fig.~5, $v_{\rm wf}$ is a 
constant, equal to $c$, independent of the frequency or details of the absorption
band.) This is determined by the boundary between the regions of zero
and non-zero disturbance (more generally, a discontinuity in the first
or higher derivative of the field) as the pulse propagates.
Mathematically, this definition of wavefront is identified with the 
characteristics of the partial differential equation governing the 
wave propagation\cite{CH}. Our problem is therefore to determine the velocity
associated with the characteristics of the wave operator derived from the
modified Maxwell equations of motion appropriate to the new effective 
action. 

Notice that a very complete and rigorous discussion of the wave equation in curved
spacetime has already been given in the monograph by Friedlander\cite{Fried}, in 
which it is proved (Theorem 3.2.1) that the characteristics are simply the 
null hypersurfaces of the spacetime manifold, in other words that the 
wavefront always propagates with the fundamental speed $c$. However, this 
discussion assumes the standard form of the (gauge-fixed) Maxwell wave 
equation (cf.~ref.\cite{Fried}, eq.(3.2.1)) and does {\it not} 
cover the modified wave equation derived from the action Eq.{\ref{eq:aa}, 
precisely because of the extra curvature couplings which lead to the 
effective metric $G_{\m\n}$ and superluminal propagation.

\subsection{Characteristics, Wavefronts and the Phase Velocity $v_{\rm ph}(\infty)$}

Instead, the key result which allows a derivation
of the wavefront velocity is derived by Leontovich\cite{Leon}.
In this paper\footnote{I am very grateful to A. Dolgov, V. Khoze and
I. Khriplovich for their help in obtaining and interpreting ref.\cite{Leon}.}, 
an elegant proof is presented for a very general set of 
PDEs that the wavefront velocity associated with the characteristics is 
identical to the $\w\rta\infty$ limit of the phase velocity, i.e.
\begin{equation}
v_{\rm wf} = \lim_{\w\rta \infty}{\w\over|\ul k|} = 
\lim_{\w\rta \infty}v_{\rm ph}(\w)
\label{eq:cb}
\end{equation}
The proof is rather formal, but is of sufficient generality to apply to our
discussion of photon propagation using the modified effective action
of section 4. We reproduce the essential details below.

The first step is to recognise that any second order PDE can be written as
a system of first order PDEs by considering the first derivatives of the field
as independent variables. Thus, if for simplicity we consider a general second 
order wave equation for a field $u(t,x)$ in one space dimension, the system of
PDEs we need to solve is
\begin{equation}
a_{ij} {\partial\phi_j\over\partial t} + b_{ij} {\partial\phi_j\over\partial x}
+ c_{ij} \phi_j ~=~0
\label{eq:cc}
\end{equation}
where $\phi_i = \{u, {\partial u\over\partial t}, {\partial u\over\partial x}\}$.

Making the `geometric optics' ansatz
\begin{equation}
\phi_i ~=~ \varphi_i \exp i(wt-kx)
\label{eq:cd}
\end{equation}
where the frequency-dependent phase velocity is $v_{\rm ph}(k) = \w(k)/k$, 
and substituting into Eq.(\ref{eq:cc}) we find
\begin{equation}
\Bigl(i\w a_{ij} - ik b_{ij} + c_{ij} \Bigr) \varphi_j ~=~ 0
\label{eq:ce}
\end{equation}
The condition for a solution,
\begin{equation}
{\rm det}\Bigl[a_{ij} v_{\rm ph}(k) - b_{ij} -{i\over k} c_{ij} \Bigr] ~=~ 0
\label{eq:cf}
\end{equation}
then determines the phase velocity.

On the other hand, we need to find the characteristics of Eq.(\ref{eq:cc}),
i.e.~curves ${\cal C}$ on which Cauchy's theorem breaks down and the evolution is not
uniquely determined by the initial data on ${\cal C}$. The derivatives of the field
may be discontinuous across the characteristics and these curves are associated with 
the wavefronts for the propagation of a sharp-fronted pulse. The corresponding light rays
are the `bicharacteristics'. (See, for example, ref.\cite{CH} chapters 5.1, 6.1 for
further discussion.) 

We therefore consider a characteristic curve ${\cal C}$ in the $(t,x)$ plane separating
regions where $\phi_i =0$ (ahead of the wavefront) from $\phi_i \ne 0$ (behind the
wavefront). At a fixed point $(t_0,x_0)$ on ${\cal C}$, the absolute derivative of 
$\phi_i$ along the curve, parametrised as $x(t)$, is just
\begin{equation}
{d\phi_i\over dt} ~=~ {\partial \phi_i\over\partial t}\Big|_0
+ {\partial \phi_i\over\partial x}\Big|_0 {dx\over dt}
\label{eq:cg}
\end{equation}
where $dx/dt = v_{\rm wf}$ gives the wavefront velocity. 
Using this to eliminate ${\partial\phi_i\over\partial t}$ from the PDE Eq.(\ref{eq:cc})
at $(t_0,x_0)$, we find
\begin{equation}
\Bigl( - a_{ij} {dx\over dt} + b_{ij}\Bigr) {\partial \phi_j\over\partial x}\Big|_0
+ a_{ij} {d\phi_j^{(0)}\over dt} + c_{ij}\phi_j^{(0)} ~=~ 0
\label{eq:ch}
\end{equation}
Now since ${\cal C}$ is a wavefront, on one side of which $\phi_i$ vanishes
identically, the second two terms above must be zero.
The condition for the remaining equation to have a solution is simply
\begin{equation}
{\rm det}\Bigl[ a_{ij} v_{\rm wf} - b_{ij} \Bigr] ~=~ 0
\label{eq:ci}
\end{equation}
which determines the wavefront velocity $v_{\rm wf}$.
The proof is now evident. Comparing Eqs.(\ref{eq:cf}) and (\ref{eq:ci}), we
clearly identify
\begin{equation}
v_{\rm wf} ~=~ v_{\rm ph}(k\rightarrow \infty)
\label{eq:cj}
\end{equation}

The wavefront velocity in a gravitational background is therefore
not given {\it a priori} by $c$. Taking vacuum polarisation into account,
there is no simple non-dispersive medium corresponding to the 
vacuum of classical Maxwell theory in which the phase velocity 
represents a true speed of propagation; for QED in curved spacetime, even
the vacuum is dispersive.
In order to discuss causality, we therefore have to extend the original
Drummond-Hathrell results for $v_{\rm ph}(\w \sim 0)$ to the high frequency
limit $v_{\rm ph}(\w\rta\infty)$, as already emphasised in their original 
work. 

A further subtle question arises if we write the standard dispersion relation
for the refractive index $n(\w)$ in the limit $\w\rta\infty$:
\begin{equation}
n(\infty) = n(0) - {2\over\pi} \int_0^\infty {d\w\over\w} \Im n(\w)
\label{eq:ck}
\end{equation}
For a conventional dispersive medium, $\Im n(\w) > 0$, which implies that
$n(\infty) < n(0)$, or equivalently $v_{\rm ph}(\infty) > v_{\rm ph}(0)$.
Evidently this is satisfied by Fig.~5. The key question though is 
whether the usual assumption of positivity of $\Im n(\w)$ holds 
in the present situation of the QED vacuum in a gravitational field.
If so, then (as already noted in ref.\cite{DH}) the superluminal
Drummond-Hathrell results for $v_{\rm ph}(0)$ would actually be {\it lower
bounds} on the all-important wavefront velocity $v_{\rm ph}(\infty)$.
However, it is not clear that positivity of $\Im n(\w)$ holds
in the gravitational context. Indeed it has been explicitly
criticised by Dolgov and Khriplovich in refs.\cite{DK,Khrip},
who point out that since gravity is an inhomogeneous medium in which beam
focusing as well as diverging can happen, a growth 
in amplitude corresponding to $\Im n(\w) < 0$  is possible. 
The possibility of $v_{\rm ph}(\infty) < v_{\rm ph}(0)$, and in particular
$v_{\rm ph}(\infty) = c$, therefore does not seem to be convincingly ruled out
by the dispersion relation Eq.({\ref{eq:ck}).

\section{Dispersion}

After these general considerations, we now return to QED in curved spacetime
and use the full effective action to study dispersion and investigate
the high-frequency limit.

\subsection{Effective Action for Photon-Gravity Interactions}

The local effective action in QED to $O(RFF)$ keeping terms of all orders in derivatives
is derived in ref.\cite{Ssix}. The result is:
\begin{equation}
\begin{array}{rcl}
&\\
&\C  = \int dx \sqrt{-g} \biggl[-{1\over4}F_{\m\n}F^{\m\n}~
+~{1\over m^2}\Bigl(D_\m F^{\m\l}~ \overrightarrow{G_0}~ D_\n F^\n{}_{\l} \\[4pt]
&~~~~~~+~\overrightarrow{G_1}~ R F_{\m\n} F^{\m\n}~ 
+~\overrightarrow{G_2}~ R_{\m\n} F^{\m\l}F^\n{}_{\l}~
+~\overrightarrow{G_3}~ R_{\m\n\l\r}F^{\m\n}F^{\l\r} \Bigr)\\
&~+~{1\over m^4}\Bigl(\overrightarrow{G_4}~ R D_\m F^{\m\l} D_\n F^\n{}_{\l}~ 
+~\overrightarrow{G_9}~ R_{\m\n\l\r} D_\s F^{\s\r}D^\l F^{\m\n}\\
&~~~~+~\overrightarrow{G_5}~ R_{\m\n} D_\l F^{\l\m}D_\r F^{\r\n}~
+~\overrightarrow{G_6}~ R_{\m\n} D^\m F^{\l\r}D^\n F_{\l\r} \\
&~~~~~~~~~~+~\overrightarrow{G_7}~ R_{\m\n} D^\m D^\n F^{\l\r} F_{\l\r}~
+~\overrightarrow{G_8}~ R_{\m\n} D^\m D^\l F_{\l\r} F^{\r\n}~\Bigr) ~~\biggr] \\
&&
\end{array}\label{eq:da}
\end{equation}
This was found by adapting a background field action valid to third order in 
generalised curvatures due to Barvinsky, Gusev, Zhytnikov and Vilkovisky\cite{BGVZone} 
(see also ref.\cite{BGVZtwo}) 
and involves re-expressing their more general result in manifestly local form by an 
appropriate choice of basis operators.

In this formula, the $\overrightarrow{G_n}$ ($n\ge 1$) are form factor functions of 
three operators:
\begin{equation}
\overrightarrow{G_n} \equiv G_n\Bigl({D_{(1)}^2\over m^2}, {D_{(2)}^2\over m^2}, 
{D_{(3)}^2\over m^2}\Bigr)
\label{eq:db}
\end{equation}
where the first entry ($D_{(1)}^2$) acts on the first following term
(the curvature), etc. $\overrightarrow{G_0}$ is similarly defined as a single variable 
function. These form factors are found using heat kernel methods and are
given by `proper time' integrals of known algebraic functions. Their 
explicit expressions can be found in ref.\cite{Ssix}. 
Evidently, Eq.(\ref{eq:da}) reduces to the Drummond-Hathrell action if we neglect all 
the higher order derivative terms.

\subsection{Dispersion and the Light Cone}

The next step is to derive the equation of motion analogous to Eq.(\ref{eq:bg})
from this generalised effective action and to apply geometric optics to find
the corresponding light cone. This requires a very careful analysis of the
relative orders of magnitudes of the various contributions to the equation
of motion arising when the factors of $D^2$ in the form factors act on the 
terms of $O(RF)$. These subtleties are explained in detail in ref.\cite{Sfive}.
The final result for the new effective light cone has the form
\begin{equation}
k^2 ~-~ {1\over m^2} F\Bigl({k.D\over m^2}\Bigr) R_{\m\l} k^\m k^\l ~+~
{1\over m^2} G\Bigl({k.D\over m^2}\Bigr) R_{\m\n\l\r} k^\m k^\l a^\n a^\r ~~=~~0
\label{eq:dc}
\end{equation} 
where $F$ and $G$ are known functions with well-understood asymptotic 
properties\cite{Sfive}. Clearly, for agreement with Eq.(\ref{eq:bh}),
we have $F(0) = 2b$, $G(0) = 8c$.

The novel feature of this new light cone condition is that $F$ and $G$ are
functions of the {\it operator} $k.D$ acting on the Ricci and Riemann
tensors.\footnote{Note that because these corrections are already of $O(\a)$,
we can freely use the usual Maxwell relations $k.D k^\n = 0$ and $k.D a^\n = 0$
in these terms; we need only consider the effect of the operator $k.D$ acting
on $R_{\m\n}$ and $R_{\m\n\l\r}$.} So although the asymptotic behaviour of $F$
and $G$ as functions is known, this information is not really useful unless 
the relevant curvatures are eigenvalues of the operator. On the positive side, 
however, $k.D$ does have a clear geometrical interpretation -- it simply describes 
the variation along a null geodesic with tangent vector $k^\m$. 

The utility of this light cone condition therefore seems to hinge on what we know
about the variations along null geodesics of the Ricci and Riemann
(or Weyl) tensors. It may therefore be useful to re-express Eq.(\ref{eq:dc}) in
Newman-Penrose form:
\begin{equation}
k^2 ~-~ {\w^2\over m^2} \tilde F\Bigl({\w \ell.D\over m^2}\Bigr) \Phi_{00} ~\pm~
{\w^2\over 2 m^2} G\Bigl({\w \ell.D\over m^2}\Bigr) (\Psi_0 + \Psi_0^*) ~~=~~0
\label{eq:dd}
\end{equation} 
where $\tilde F = 2F + G$.

Unfortunately, we have been unable to find any results in the relativity literature
for $\ell.D \Phi_{00}$ and $\ell.D \Psi_0$ which are valid in a general spacetime. 
In particular, this is {\it not} one of the combinations that are constrained by 
the Bianchi identities in Newman-Penrose form (as displayed for example in 
ref.\cite{Chand}, chapter 1, Eq.(321)). To try to build some intuition,
we have therefore looked at particular cases. The most interesting is
the example of photon propagation in the Bondi-Sachs metric\cite{Bondi,Sachs} 
which we recently studied in detail\cite{Sfour}.
  
The Bondi-Sachs metric describes the gravitational radiation from an
isolated source. The metric is
\begin{equation}
ds^2 = -W du^2 - 2 e^{2\b} du dr + r^2 h_{ij}(dx^i - U^i du)
(dx^j - U^j du)
\label{eq:de}
\end{equation}
where
\begin{equation}
h_{ij}dx^i dx^j = {1\over2}(e^{2\c} + e^{2\d}) d\theta^2
+ 2 \sinh(\c - \d) \sin\theta d\theta d\phi
+ {1\over2}(e^{-2\c} + e^{-2\d}) \sin^2\theta d\phi^2
\label{eq:df}
\end{equation}
The metric is valid in the vicinity of future null infinity ${\cal I}^+$.
The family of hypersurfaces $u = const$ are null, i.e. $g^{\m\n}
\pl_\m u \pl_\n u = 0$. Their normal vector $\ell_\m$ satisfies
\begin{equation}
\ell_\m = \pl_\m u ~~~~~~~~~~~~~\Rightarrow~~~~~
\ell^2 = 0, ~~~~~~~~\ell^\m D_\mu \ell^\n = 0
\label{eq:dg}
\end{equation}
The curves with tangent vector $\ell^\m$ are therefore
null geodesics; the coordinate $r$ is a radial parameter along these rays  
and is identified as the luminosity distance. 
The six independent functions  $W,\b,\c,\d,U^i$
characterising the metric have expansions in 
${1\over r}$ in the asymptotic region near ${\cal I}^+$, the coefficients of
which describe the various features of the gravitational radiation.

In the low frequency limit, the light cone is given by the simple
formula Eq.({\ref{eq:ad}) with $\Phi_{00} = 0$. 
The velocity shift is quite different for the case of outgoing
and incoming photons\cite{Sfour}. For outgoing photons, $k^\m = \ell^\m$,
and the light cone is
\begin{equation}
k^2 ~~=~~\pm~{4c\w^2\over m^2}~\Bigl(\Psi_0 + \Psi_0^*\Bigr) ~~\sim ~~
O\Bigl({1\over r^5}\Bigr)
\label{eq:dh}
\end{equation}
while for incoming photons, $k^\m = n^\m$,
\begin{equation}
k^2 ~~=~~\pm~{4c\w^2\over m^2}~\Bigl(\Psi_4 + \Psi_4^*\Bigr) ~~\sim ~~
O\Bigl({1\over r}\Bigr)
\label{eq:di}
\end{equation}

Now, it is a special feature of the Bondi-Sachs spacetime 
that the absolute derivatives of each of the Weyl
scalars $\Psi_0, \ldots, \Psi_4$ along the ray direction $\ell^\m$
vanishes, i.e.~$\Psi_0, \ldots, \Psi_4$ are parallel transported along 
the rays\cite{Sachs,Inverno}. In this case, therefore, we have: 
\begin{equation}
\ell\cdot D ~\Psi_0 ~=~0 ~~~~~~~~~~~~~~~~~~~~~~~~~~~~
\ell\cdot D ~\Psi_4 ~=~0
\label{eq:dj}
\end{equation}
but there is no equivalent simple result for either $n\cdot D ~\Psi_4$
or $n\cdot D ~\Psi_0$.

Although it is just a special case, Eq.(\ref{eq:dj}) nevertheless 
leads to a remarkable conclusion. The full light cone condition 
Eq.(\ref{eq:dd}) applied to outgoing photons in the Bondi-Sachs spacetime now 
reduces to
\begin{equation}
k^2 ~\pm~ {\w^2\over 2 m^2}~ G\bigl(0\bigr)
\Bigl(\Psi_0 + \Psi_0^*\Bigr) ~~=~~ 0
\label{eq:dk}
\end{equation}
since $\ell\cdot D \Psi_0 = 0$. In other words, the low-frequency 
Drummond-Hathrell prediction of a superluminal phase velocity $v_{\rm ph}(0)$
is {\it exact} for all frequencies. There is no dispersion, and the 
wavefront velocity $v_{\rm ph}(\infty)$ is indeed superluminal. 

This is potentially a very important result. Based on the improved effective action 
Eq.(\ref{eq:da}), we have shown there is at least one example in which the
wavefront truly propagates with superluminal velocity. Quantum effects
have indeed shifted the light cone into the geometrically spacelike region.

\subsection{Non-Perturbative Effective Action and High-Frequency Propagation}

Unfortunately, there is one final twist to the story which could invalidate
the above conclusion. If instead of a gravitational field, we consider photon 
propagation in a constant background magnetic field, we find the following 
birefringent modification to the light cones:\cite{TEone,TEtwo,Sfive}
\begin{equation}
k^2~~-~~ {\a\over2\pi} \Bigl({eB\w\over m^2}\Bigr)^2 ~
\int_{-1}^1 du \int_0^\infty ds ~s ~
N_{\parallel,\perp}(u,z)~ e^{-is\bigl(1 + s^2 \W^2 P(u,z)\bigr)}  ~~=~~ 0
\label{eq:dl}
\end{equation}
where $z = {eB\over m^2}s$ and $\W = {eB\over m^2}{\w\over m}$.
Exact expressions are known for the functions $N$ and $P$ as well as their
weak-field expansions.

Now, in the weak-field, low-frequency regime, $\Omega \ll 1$ and we can simply 
expand the exponential to leading order in $\Omega^2 P$. The low-frequency limit 
of the phase velocity for the two polarisations is the well-known 
result\cite{Sone,Adler} 
\begin{equation}
v_{\rm ph}(0) ~~=~~ 1 - {\a\over4\pi} \Bigl({eB\over m^2}\Bigr)^2~
\Bigl[{14\over45}_\parallel, {8\over45}_\perp \Bigr]
\label{eq:dm}
\end{equation} 
and is in each case subluminal. At high frequencies, however, 
Eq.(\ref{eq:dl}) is dominated by the rapidly-varying phase factor 
$\exp(-i s^3 \Omega^2 P)$ and 
the correction to the phase velocity tends to zero (from a superluminal value)
with a non-analytic $\Omega^{-{4/3}}$ behaviour:
\begin{equation}
v_{\rm ph}(\w\rightarrow \infty) ~~=~~ 1 + {\a\over4\pi} \Bigl({eB\over m^2}\Bigr)^2~
\bigl[{c}_\parallel, {c}_\perp \bigr]~ \W^{-{4\over3}}
\label{eq:dn}
\end{equation} 
where ${c}_\parallel, {c}_\perp$ are known positive constants\cite{Sfive}.
The phase velocity therefore has the standard form for a dispersive medium
illustrated in Fig.~5. In particular, the wavefront velocity $v_{\rm wf}
= v_{\rm ph}(\infty) = c$.

The important lesson for the gravitational case is this. If we had simply used
an effective action for QED in a background electromagnetic field keeping
terms up to $O(F^4)$ in the field strength and to all orders in derivatives,
generalising the Euler-Heisenberg Lagrangian, we would have accurately found 
the leading low-frequency dependence of $v_{\rm ph}(\w)$
but would have completely missed the non-analytic high-frequency behaviour.
This arises from terms in the full effective action which are {\it 
non-perturbative} in the background field and give rise to the phase
factor in Eq.(\ref{eq:dl}).

If the gravitational case is similar, this would imply that the modified
light cone can be written heuristically as
\begin{equation}
k^2 ~+~ {\a\over\pi} \int_0^\infty ds~{\cal N}(s,R)~
e^{-is\bigl(1 + s^2\W^2 {\cal P}(s,R)\bigr)}~~=~~0
\label{eq:do}
\end{equation}
where both ${\cal N}$ and ${\cal P}$ can be expanded in powers of curvature,
and derivatives of curvature, presumably associated with factors of $\w$ as in 
the last section. The frequency
dependent factor $\W$ would be $\W \sim {R\over m^2}{\w\over m} \sim
O({\l_c^3\over\l L^2})$, where `$R$' denotes some generic curvature component 
and $L$ is the typical curvature scale. If this is true, then an expansion of 
the effective action to $O(RFF)$, even including higher derivatives, 
would not be sufficient to reproduce the full, non-perturbative contribution 
$\exp(-is^3\Omega^2{\cal P})$. The Drummond-Hathrell action would correspond 
to the leading order term in the expansion of Eq.(\ref{eq:do}) in powers of 
${R\over m^2}$ neglecting derivatives, while our improved effective action
of section 4.1 sums up all orders in derivatives while retaining the restriction 
to leading order in curvature.  

The omission of the non-perturbative contribution would be justified only in the 
limit of small $\W$, i.e.~for ${\l_c^3\over\l L^2} \ll 1$. Neglecting this 
therefore prevents us from accessing the genuinely high frequency limit
$\l \rta 0$ needed to find the asymptotic limit $v_{\rm ph}(\infty)$ of the phase 
velocity. Moreover, assuming Eq.(\ref{eq:do}) is indeed on the right lines, 
it also seems inevitable that for high frequencies (large $\W$) the rapid phase
variation in the exponent will drive the entire heat kernel integral to zero,
ensuring the wavefront velocity $v_{\rm wf} = c$.

\subsection{Outlook}

At present, it is not clear how to make further progress. The quantum field theoretic
calculation required to find such non-perturbative contributions to the effective 
action and confirm an $\exp(-is^3\Omega^2{\cal P})$ structure in Eq.(\ref{eq:do})
appears difficult, although some technical progress in this area has been
made recently in ref.\cite{BM} and work in progress\cite{Gusev}.
One of the main difficulties is that since a superluminal effect requires 
some anisotropy in the curvature, it is not sufficient just to consider 
constant curvature spacetimes. (Recall that the Ricci scalar term in the 
effective action Eq.(\ref{eq:aa}) does not contribute to the modified light 
cone Eq.(\ref{eq:ab}).) A possible approach to this problem, which would help
to control the plethora of indices associated with the curvatures, might be 
to reformulate the heat kernel calculations directly in the Newman-Penrose
basis. On the other hand, perhaps a less ambitious goal would be to try to
determine just the asymptotic form of the non-perturbative contribution in the 
$\Omega\rta\infty$ limit.

A final resolution of the dispersion problem for QED in curved spacetime has 
therefore still to be found. If the perturbative expansion of the effective
action to $O(RFF)$ is sufficient, then as we have seen there exist at least
some examples where the wavefront velocity is really superluminal.
In this case, all the issues concerning causality discussed in section 2
would apply to QED. Perhaps more likely, however, is the scenario described
above, where the high-frequency dispersion is driven by 
non-perturbative contributions to the effective action such that the wavefront
velocity remains precisely $c$. It would then be interesting to see exactly
how the phase velocity behaves as a function of $\w$ (c.f.~Fig.~5) and whether the
explanation advanced in section 3.2 for the non-validity of the standard
refractive index dispersion relation Eq.(\ref{eq:ck}) is correct.

Finally, of course, even if it does turn out that $v_{\rm wf} = c$ for QED itself,
the discussion of causality in this paper may still be relevant to photon
propagation in more speculative theories, including bimetric theories
of gravity\cite{Drum}, string-inspired non-linear electrodynamics\cite{Gibb},
Lorentz and CPT violating effective Lagrangians\cite{Kost} and 
non-commutative gauge theories.

\section*{Acknowledgments}

This work is supported in part by PPARC grant PP/G/O/2000/00448.

\end{document}